# Design of iMacros-based Data Crawler and the Behavioral Analysis of Facebook Users

**Mudasir Ahmad Wani**
*Research laboratory Department Computer Science Faculty of Natural Science Jamia Millia Islamia (A Central University) New Delhi, India*
mudassir148081@st.jmi.ac.in

**Nancy Agarwal**
*Research laboratory Department Computer Science Faculty of Natural Science Jamia Millia Islamia (A Central University) New Delhi, India*
nansy148081@st.jmi.ac.in

**Suraiya Jabin**
*Department of Computer Science Faculty of Natural Science Jamia Millia Islamia (A Central University) New Delhi, India*
sjabin@jmi.ac.in

**Syed Zeesahn Hussain**
*Department of Computer Science Faculty of Natural Science Jamia Millia Islamia (A Central University) New Delhi, India*
szhussain@jmi.ac.in

**Abstract**

Obtaining the desired dataset is still a prime challenge faced by researchers while analyzing Online Social Network (OSN) sites. Application Programming Interfaces (APIs) provided by OSN service providers for retrieving data impose several unavoidable restrictions which make it difficult to get a desirable dataset. In this paper, we present an iMacros technology-based data crawler called IMcrawler, capable of collecting every piece of information which is accessible through a browser from the Facebook website within the legal framework which permits access to publicly shared user content on OSNs. The proposed crawler addresses most of the challenges allied with web data extraction approaches and most of the APIs provided by OSN service providers. Two broad sections have been extracted from Facebook user profiles, namely, Personal Information and Wall Activities. The present work is the first attempt towards providing the detailed description of crawler design for the Facebook website.

## Introduction

Online Social Network (OSN) is a web application which focuses on the social life of netizens and provides them an excellent platform to build a network of relationships and share their social life. OSNs such as Facebook, LinkedIn and Twitter are increasingly becoming popular among internet users and turning to be an essential medium for people's daily activities. Since their structure significantly reflects the real-life communities and they contain a huge amount of user's personal and social information, they are of scientific importance to the researchers and disciplines in different domains including marketing, sociology, politics etc. Marketers study OSNs to design a viral marketing strategy (Staab et al., 2005), sociologists use them to study the human behavior (Christakis et al., 2013) and politicians use them to empower their political campaigns (Vergeer et al., 2013).

For conducting any kind of analysis, sufficient amount of data is the preliminary requirement (Wani et al., 2018). OSNs like Facebook contain billions of user profiles and the service providers ensure that their data is protected which makes the process of data collection very challenging for researchers. OSN datasets are not publicly available because of privacy reasons and since the data is enormous, collecting it manually makes the task complex and time-consuming. However, most popular social networking sites like Facebook, Twitter, Flickr, etc. provide methods for retrieving information from the network through their own well defined Application Programming Interfaces (APIs) like Graph-API[1], REST-API[2] etc., but these APIs are associated with several unavoidable constraints such as data request rate restrictions, selective data access, etc. Web scrapping provides an alternative solution by automatically extracting the information from the web pages in a systematic manner. Although it can solve the problem of data collection to a great extent but writing a scrapper is itself a challenging task. Generally, social media platforms including Facebook have inbuilt bot detection mechanism (Stein et al., 2011) which can recognize an automated activity on their platforms and, therefore data collection by means of software programs may lead to the suspension of the user account which is being used for data extraction. The feature of dynamic loading of content via various web technologies (e.g. JavaScript and Ajax) further complicates the task since this information is not available in the source code of a web page. Moreover, the call to dynamic content on the web page is mostly triggered by user interactions with the page which implies that there should be a mechanism to automate these interactions in order to load this dynamic content into the parent HTML document. Therefore, there is a need for a tool which is capable of circumventing the API constraints and can overcome the data scrapping barriers.

---
[1] https://developers.facebook.com/docs/graph-api
[2] https://dev.twitter.com/rest/public

The paper focuses on the designing of data crawler, IMcrawler for a Facebook network that addresses the above-discussed challenges and assists the end user in extracting the data in an efficient and convenient manner. Facebook is one of the topmost networking sites and has the most complex privacy policy structure. It's API can be used to extract data from only those users who are already registered with the application. Unlike Twitter API, the Facebook API has to explicitly ask for permissions from its members to access their data. Users' privacy settings and privileges granted to the application will decide the data that can be scrapped from their profiles. Facebook has also imposed constraints on the maximum request rate to limit the amount of data being scrapped. The crawler proposed in this paper does not face any of these hurdles and is able to extract every information of any amount which appears on the user profile. It is also capable of interacting with Timeline to load the dynamic content. The complete framework of the data collection is also described with step wise processes followed from crawling the network to obtain the featured dataset in a useful format.

The crawler has been designed to extract two broad sections viz profile information (profile features) and wall activities (post features), particularly from the friends of a profile. Profile information consists of details provided by the users about themselves and wall activities consist of actions performed by the users on the Timeline. The collected information is divided into two datasets. The first dataset represents the attributes of a profile and the second dataset holds the post information. The remaining article is structured as follows: Initially, in section 2, the work relevant to data extraction from social networks and analysis on the information revealed by users has been discussed. Section 3 describes the novel contribution of the proposed work. The design of the IMcrawler and it's working have been explained in section 4. Finally, the section 5 concludes the overall work and provides future directions.

## 2. Related Work

Given the massive amount of online user data, its extraction is the key challenge for the researchers. One of the most recent studies (Zhou, D. et al., 2018) presented a detailed survey on the existing data collection methods, mechanisms and architectures along with some open challenges associated with the process of data collection. The authors (Ferrara et al., 2014) in one more study has presented a detailed discussion on web data extraction techniques along with the application domains in which they are applied. In general, APIs and HTML scrapping are the two popular ways to retrieve data from OSNs. Although APIs provide well-organized data but with several restrictions. The HTML scrapping techniques offer an alternative solution that can circumvent the limitations imposed by APIs but at the price of technical complexities. The paper (Maynard et al., 2017) presented a semantic-based framework for collecting the social media data using APIs and analyzing it by the open source tools provided by GATE family[3]. The authors in (Mislove et al. 2007) have analyzed the structure of several popular OSNs such as YouTube, Flicker, LiveJournal, etc. The study has used the APIs provided by these OSNs and HTML scrapping technique in order to extract the required data. Data extraction schemes significantly depend on the policy of online social networks as in (Alim et al., 2009), the authors have presented the process of extracting personal attributes and the list of top friends from MySpace social network without being its member. A study (Rungsawang et al., 2005) highlights the pros and cons of the existing data crawlers and proposed an algorithm to resolve the issues associated with previous data crawlers. MySpace provides a rich source of data to the nonmembers as well, whereas networks like Facebook, Friendster etc. expose either no content or minimal content to the external users. Several studies have been carried out on the Facebook network exclusively since it is one of the most popular online social networking websites and has the most complex privacy mechanism (Dwyer et al., 2007, Acquisti et al., 2006). Netviz (Rieder, 2013) is a Facebook application designed to help researchers in collecting features of a profile including personal networks, groups, and pages. However, just like any other API, the working of Netvizz application is also limited by the permission and privacy model of the Facebook service. First, it requires logged in Facebook account. Second, it explicitly asks the user for the permission to access their different data and third, the user can further restrict the data availability to the application by their privacy settings. In (Wong et al., 2014), the authors have implemented an HTML-based crawler by using PhantomJS, a headless browser[4], to extract the friend's network of Facebook users of a specific region in Macau. In addition, authors have also discussed the technical challenges and their viable solutions while designing the data extractor for OSNs.

---
[3] https://gate.ac.uk/family

OSN sites offer a great medium to its users to share varied amount of social as well as personal details with either all the members of the network or a specific community of users (Javed, M. A. et al., 2018). A number of studies have been conducted to explore the pattern of information disclosed by the users of these websites. The paper (Nosko et al., 2010) examines the kind of information that has been disclosed most on the Facebook profiles and what type of people are likely to disclose it more. The authors categorized the revealed information into three groups viz identity information (schools, jobs, etc), sensitive information (email, profile picture, etc.) and stigmatizing information (religious views, political views, etc). The authors observed that people who mentioned details about their gender, relationship status, and age have revealed more information about themselves in comparison to the users who did not provide the said details. Moreover, age and relationship status were found to be the salient features in predicting the disclosure of information. Another study (Al-Saggaf et al., 2014) has shown that the users who feel lonely, specifically females, reveal information about themselves that might encourage other users to contact them such as relationship status, address, etc. However, exposing too much on OSNs can cause serious problem to the users due to the severe risks associated with the revealed information such as identity theft, cyber stalking, etc. (Bilge et al., 2009). Given the enormous use and potential danger associated with the information disclosed, it is important for the users to understand the privacy settings provided by social sites so as to protect their information from getting in wrong hands. In (Gross et al., 2005), authors have conducted a study on the Facebook network of a university campus. They investigated the patterns of personal information disclosed by its students and usage of privacy features offered by the site for limiting the visibility of the content. It has been found that while the personal details are generally revealed, privacy features are hardly concerned.

Unlike the web, which is primarily centered on content, online networking sites are concerned with the user life. These sites allow users to publish information about them, connect to each other, share content, and disseminate information and so on. Hence all these activities can be analyzed to observe a variety of trends on the network. In a paper (Viswanath et al., 2009), the authors have extracted the data from New Orleans regional network of Facebook for extracting friendship links and wall activities and studied the evolution of the relationship between users on the basis of their interaction on the wall. A study (Wilson et al., 2009) has used Python programming language to obtain data from the Facebook network and proposed interaction graph to represent real social relationships. They showed that the active social links of the users are much lower as compared to their total friends in the network.

## 3. Contribution of the Proposed Work

---

[4] http://phantomjs.org

Although a number of researchers (Catanese et al., 2011, wong et al., 2014) have crawled the OSN sites, but the least attention has been paid towards providing the technical details of the data collection process and its complexities. The presented IMcrawler positively overcomes most of the demerits coupled with available APIs and the challenges with HTML scrapping. The paper discusses the design and implementation of data collection approach in a well defined, comprehensive and systematic manner. It also introduces a new technique called iMacros for the collection of user-related information from OSN sites. The present work will prove to be fruitful for the upcoming researchers by providing them deep insight into the data collection mechanism and enable them to overcome the challenges faced while data collection. Furthermore, the paper also discusses the several statistical observations and findings from the collected datasets pertaining to user's behavioral analysis. We sampled around 10,000 profiles based on the four metropolitan cities (Bangalore, Delhi, Mumbai, Pune) of India, filtered using current city information from the whole data collected through seed profiles on the Facebook network. Our analysis is divided into two areas: *personal information disclosure* and user *wall activities*. The contribution of the work is briefly discussed as follows:

- An iMacros technology-based data crawler for the Facebook network has been designed and implemented.
- The whole data collection procedure, its technical aspects, and complexities have been explained in detail.

## 4. Design of IMcrawler

Online Social networking sites contain rich amount of information about its users which is mainly stored in a semi-structured format as HTML documents. Before performing any analysis, these huge amounts of data present on any OSN needs to be extracted in a systematic and automated manner. Web extraction tools are one of the ways to efficiently collect data from a website with least human effort. There are a number of data extraction approaches such as Natural Language Processing (NLP)- based approaches, Ontology-based approaches, HTML-aware approaches, etc. (Laender et al., 2002). Here, in this paper, we are discussing the implementation of an iMacros technology-based data crawler for the Facebook social network. iMacros is a browser extension specifically designed to automate web interactions and extract data from a website[5]. It allows scrapping every piece of data from a web page which is accessible thorough a browser. iMacros add-on is available for almost all the prevalent Web Browsers including Mozilla Firefox, Google Chrome and Internet Explorer. However, the working may slightly vary from browser to browser. iMacros uses Document Object Model (DOM) tree (Gupta et al., 2003) to identify the data to be scrapped. DOM provides the structured representation of a web page where nodes denote the HTML tags and tree hierarchy represents the organization of the nested elements. Generally, dynamic applications bind the information from the databases with predefined web templates as per the user request. It implies that a large section of HTML structure of a web page remains almost static for different profiles, and only the content of the pages differs. These static elements in a page assist in extracting varied content from multiple profiles.

In order to extract the data, we need to crawl through the targeted OSN. Basically, an OSN is a graph where each profile is considered as a node and friend-links represent edges between the nodes. However, crawling complete graph is practically not possible for a network like Facebook which has billions of profiles (nodes). Hence, a relatively small but representative sub-graph as per the problem to be solved can be considered to serve the purpose. Graph traversal techniques and randomly selecting profiles based on a certain condition are the two well-known techniques used to gather the data sample that describes the network. In this paper, we use Breadth First Search (BFS) as the graph traversal technique because it is easy to implement and has been extensively employed by different researchers for crawling profiles (Gjoka et al., 2010). The BFS algorithm starts from a root profile, explores and discovers its friends and then moves to the next level.

In order to speed up the extraction process, we run iMacros scripts on different machines (we call them crawler agents) and remotely access and control these independent machines with a centralized host machine as shown in Figure 1. A study (Chau et al., 2007) has presented a parallel framework to collect data from online auction websites. In order to further shorten the time for data collection, we created multiple user sessions on the agent machines to run multiple scripts in parallel.

---
[5] http://wiki.imacros.net/Tutorials

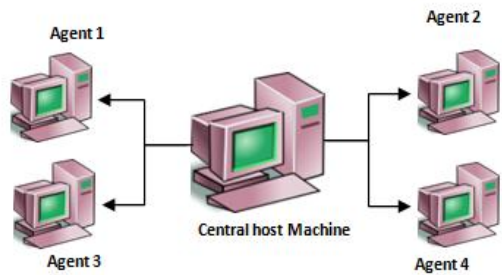

**Figure 1:** Parallel Data Collection Approach

Crawlers are designed based on the problem under consideration. More extraction of attributes needs more time and technologies. Therefore, one can design an optimized crawler by knowing in advance about the data of interest. As far as our data of interest is concerned, we have extracted most of the sections of a Facebook profile including *basic information, family, and relationships, places lived, pages liked, groups joined, post details, post emotions, post shares*, etc. and out of which we are interested in two main categories of attributes namely personal attributes and post attributes. Personal attributes include the number of *friends, birthday, e-mail address, phone number, family members, relationship status, gender, home town, current city*, etc. While the post attributes include the type of the post (*text, photo, video, etc*.), tags in a post, etc.

The complete algorithm of the IMcrawler is shown in Figure 2. Before starting the extraction process we need to select the seed profiles (registered user accounts) which act as an entry point for the crawler. The proposed crawling approach uses a set of seed accounts which have huge friend network and whose log-in credentials have been collected from their owners. This enables us to extract the friend-specific information as well. These seed profiles can be one or many, mainly depends upon the amount of data to be required. We manually selected the seed profiles from our friend lists based on the number of friends. We distributed the seed profiles among distinct crawling agents. Every agent extracts the friend links of a specified seed profile and visits each friend link one by one to extract the information.

As shown in the pseudocode, the CRAWL procedure requires two arguments: *seedFile* (contains seed profiles and their credentials) and *configFile* (contains the values of the several parameters to be used during crawling the data). Description of parameters in the configuration file is given in Table 1. The two attributes namely, *reextractLinksFile* and *seedProfile*, are basically used for re-collecting the data of the profiles whose data has not been extracted as expected. The CRAWL procedure reads the parameters from the configuration file into the *conf* object. If specific *seedProfile* is not given, the data is extracted from all the friend links for each *seedProfile* mentioned in the *seedFile*, otherwise, data is collected from the friend links mentioned in the *reextractLinksFile* by using the user account indicated by parameter *seedProfile*. The extracted data is saved to file specified by the *outputFile* parameter. Basically, extraction process requires three things. First, a source file that contains the URL of target web pages (*friendLinksFile* and *reextractLinksFile*). Second, it requires the HTML tag identifier to locate the data to be extracted from a web page. And third, a file to store the extracted data (*outputFile*).

**Table 1: Description of parameters in the configuration file**

| Parameters | Value | Description |
| --- | --- | --- |
| friendLinks | Path to the file | Contains the path to the file which contains the friend links extracted from the seed profiles. |
| outputFile | Path to the file | Contains the path to the file where the extracted data is to be stored. |
| totalPost | Numeric | Specifies the number of posts to be extracted from each user profile. |
| reextractLinksFile | Path to the file | Contains the path to the file which contains the friend links for which data is to be re-extracted. |
| seedProfile | Numeric | Value that represents the specific seed account to be used to extract the data. |

```
Algorithm 1 : IMcrawler
Require: seedFile and configFile
 1: procedure CRAWL(seedFile,ConfigFile)
 2:     conf ←read parameters from configFile
 3:     if !conf.seedProfile then
 4:         for all seedProfile in seedFile do
 5:             profileCredentials ←read seedProfile credentials from seedFile
 6:             login to profile using profileCredentials
 7:             visit friends option on timeline
 8:             extract friend links
 9:             save friend links to the file config.friendListFile
10:             EXTRACTPERSONALATTRIBUTES(conf.friendLinksFile,conf.outputFile)
11:             EXTRACTPOSTATTRIBUTES(conf.friendLinksFile,conf.totalPost,conf.outputFile)
12:             logout from profile
13:         end for
14:     else
15:         profileCredentials ←read conf.seedProfile credentials from seedFile
16:         login to profile using profileCredentials
17:         EXTRACTPERSONALATTRIBUTES(conf.reextractLinksFile,conf.outputFile)
18:         EXTRACTPOSTATTRIBUTES(conf.reextractLinksFile,conf.totalPost,conf.outputFile)
19:         logout from profile
20:     end if
21: end procedure
22: procedure EXTRACTPERSONALATTRIBUTES(friendListFile,outputFile)
23:     for all UserLink in friendListFile do
24:         visit about option on timeline
25:         extract basic − information − section
26:         extract places − lived
27:         extract family − and − relationship
28:         extract number − of − friends
29:         extract pages − liked
30:         extract groups − joined
31:         save extracted data to the file outputFile
32:     end for
33: end procedure
34: procedure EXTRACTPOSTATTRIBUTES(friendListFile,totalPosts,outputFile)
35:     for all UserLink in friendListFile do
36:         for i ← 1, totalPosts do
37:             extract post − title
38:             extract post − content
39:             extract post − date
40:             extract post − time
41:             extract post − comments
42:             extract post − emotions
43:             extract post − shares
44:             extract post − views
45:             extract post − reactions
46:             save extracted data to the file outputFile
47:         end for
48:     end for
49: end procedure
```

**Figure 2:** Algorithm for implementation of IMcrawler

The presented crawler is not limited to above mentioned attributes only. Researchers can extend this data crawler to extract the features of their interest or as per the problem under consideration. The complete framework of the IMcrawler is shown in Figure 3 that describes the step wise processes from crawling the Facebook network to obtain a sampled dataset in the required format. The whole process is divided into 5 modules. The first module deals with the setting the parameters in the configuration file for the execution of crawler program such as reference to the input files, output file, etc. In the second module, the execution of crawler program takes place to collect the data from the network. The output of the crawler program is a raw file which needs to be converted into a structured format before verifying it and this conversion process is done in step 3 (in our case the .csv file has been converted into excel sheet ). Step 4 applies several verification tests to ensure whether the data is collected as expected. The extracted data in the files have been verified manually as well. There may be the case when junk data is collected for some profiles due to technical problems like a slow internet connection or machine overload. The data is re-collected by specifying those profiles and the seed account in the *reextractLinksFile* and *seedProfile* respectively. The *reextractLinksFile* is actually a parameter which specifies the file where the links for which the data need to be re-extracted are stored and *seedProfile* parameter indicates the seed profile in the *seedFile* that will be used for re-extraction. Finally, step 5 filters the information from the markup tags in the verified file and stores into the database. It may also be possible that seed profiles have mutual friends among them which lead to the extraction of some profile data multiple times. The primary key concept of database helps to resolve it by preventing the storing of same data multiple times.

Although we designed and implemented a data crawler for the Facebook network which resolves the technical complexities and overcomes the challenges reported by most of the existing studies but it may need little modifications with the changing requirements in the research. Presently, the proposed crawler extracts only the specific piece of information (features) from a Facebook account such as wall activity and profile features. However, the other attributes such as profile picture, replies to comments, etc. can be obtained by simple modifications in the algorithms. Moreover, the crawler is designed by keeping in mind the architecture of Facebook social networking website only, therefore, in order to extract the data from other OSN sites, the algorithm needs to be redesigned as per an OSN under consideration.

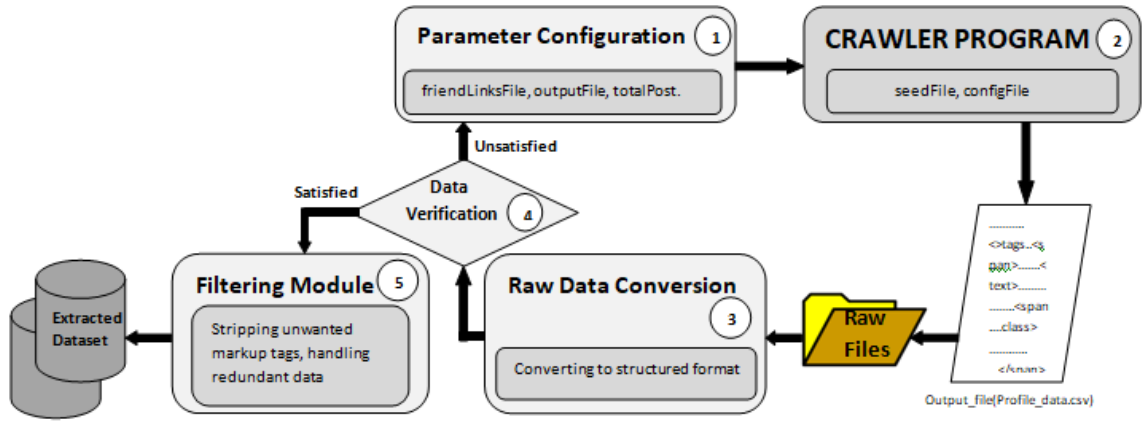

**Figure 3:** Data Collection Framework

In this paper, we have designed and implemented a data crawler for the Facebook network called *IMcrawler* with the aim to resolve the technical complexities and overcome the challenges reported by most of the existing studies. As we have observed, the IMcrawler is flexible, fast and easy to implement. The APIs such as Facebook Graph API (Weaver and Tarjan, 2013) and API-based crawlers such as (Ortigosa et al., 2014), Norconex (Norconex, 2017), Netviz (Rieder, 2013), etc. provide well-structured data but on the other hand, they are limited in terms of which data, how much data, and how often data can be retrieved. IMcrawler offers a more flexible solution by incorporating HTML scarping technique to fetch the information from the website. Different studies (Ahmed et al., 2013) have employed HTML parser package (Gupta et al., 2003) to overcome the challenges faced with Graph API but unfortunately they were able to extract the only information which is visible on the source code of an HTML page and cannot extract the dynamically loaded HTML content (information rendered by AJAX call). Our proposed crawler is sound enough to extract the dynamically loaded content. From the extraction speed point of view, we tested IMcrawler to extract the friend list of users who have more than 500 friends in their friend list and recorded the average extraction time as 10 seconds which is very less than the time reported in the literature (Wong et al., 2014). Besides, Selenium (SeleniumHQ, 2018) -a web browser automation tool has been used by various studies (Viswanath et al., 2014), (Chaabane et al., 2014) to scrap the dynamic data from the Facebook network. But it works with server-side languages such as python and java which slows down the extraction process, whereas iMacros-based crawlers operate as a browser extension and do not face any such issue.

## 5. Conclusion and Future Scope

As OSNs are a huge source of information about people, analyzing social networks may result in interesting findings and useful knowledge. But unfortunately, researchers are facing a number of challenges while analyzing the social networks because of the deficiency of ground truth data. Although OSN service providers have provided APIs to extract the data but untowardly with several unavoidable restrictions which make it difficult to get a problem specific dataset. In this paper, we have designed and implemented an iMacros technology-based data crawler called IMcrawler for Facebook network and conducted behavioral analysis on the extracted user data. IMcrawler is independent of any API and overcomes most of the challenges faced by the current researchers while extracting the data from the OSN sites. For the proposed work, we have filtered more than 10,000 profiles from the extracted data based on their current city information. We analyzed the collected data to draw several behavioral aspects about the users based on the personal information disclosed and the prevalent wall activities performed by them.

The scope of presented data crawler is not limited to the attributes discussed in this paper. Researchers will be able to conveniently extend the IMcrawler as per their requirements. Although there are some extracted attributes that have not been used in the present study but they play a significant role in our upcoming research work. Furthermore, the collected data can be analyzed from different perspectives like sentiment analysis of user posts, community detection, identification of anomalous user behavior, etc.